\documentclass[sigconf]{acmart}

\usepackage{algorithm}
\usepackage{algpseudocode}
\usepackage{multirow}
\usepackage{subcaption}
\usepackage{cleveref}
\algdef{S}[FOR]{ForEach}[1]{\algorithmicforeach\ #1\ \algorithmicdo}

\copyrightyear{2022} 
\acmYear{2022} 
\setcopyright{acmcopyright}
\acmConference[CIKM '22]{Proceedings of the 31st ACM International Conference on Information and Knowledge Management}{October 17--21, 2022}{Atlanta, GA, USA}
\acmBooktitle{Proceedings of the 31st ACM International Conference on Information and Knowledge Management (CIKM '22), October 17--21, 2022, Atlanta, GA, USA}
\acmPrice{15.00}
\acmDOI{10.1145/3511808.3557065}
\acmISBN{978-1-4503-9236-5/22/10}
\begin{document}

\title{Real-time Short Video Recommendation on Mobile Devices}

\author{Xudong Gong}
\affiliation{
    \institution{Kuaishou Inc.}
    \city{Beijing}
    \country{China}
}
\email{gongxudong@kuaishou.com}
\orcid{0000-0002-6086-7875}
\author{Qinlin Feng}
\affiliation{
    \institution{Kuaishou Inc.}
    \city{Beijing}
    \country{China}
}
\email{fengqinlin@kuaishou.com}
\author{Yuan Zhang}
\affiliation{
    \institution{Kuaishou Inc.}
    \city{Beijing}
    \country{China}
}
\email{zhangyuan13@kuaishou.com}
\author{Jiangling Qin}
\affiliation{
    \institution{Kuaishou Inc.}
    \city{Beijing}
    \country{China}
}
\email{qinjiangling@kuaishou.com}
\author{Weijie Ding}
\affiliation{
    \institution{Kuaishou Inc.}
    \city{Beijing}
    \country{China}
}
\email{dingweijie@kuaishou.com}
\author{Biao Li}
\affiliation{
    \institution{Kuaishou Inc.}
    \city{Beijing}
    \country{China}
}
\email{libiao@kuaishou.com}
\author{Peng Jiang}
\affiliation{
    \institution{Kuaishou Inc.}
    \city{Beijing}
    \country{China}
}
\email{jiangpeng@kuaishou.com}
\author{Kun Gai}
\affiliation{
    \institution{Unaffiliated}
    \city{Beijing}
    \country{China}
}
\email{gai.kun@qq.com}

\renewcommand{\shortauthors}{Xudong Gong et al.}

\begin{abstract}
Short video applications have attracted billions of users in recent years, fulfilling their various needs with diverse content. Users usually watch short videos on many topics on mobile devices in a short period of time, and give explicit or implicit feedback very quickly to the short videos they watch. The recommender system needs to perceive users' preferences in real-time in order to satisfy their changing interests.
Traditionally, recommender systems deployed at server side return a ranked list of videos for each request from client. Thus it cannot adjust the recommendation results according to the user's real-time feedback before the next request. Due to client-server transmitting latency, it is also unable to make immediate use of users' real-time feedback. However, as users continue to watch videos and feedback, the changing context leads the ranking of the server-side recommendation system inaccurate.
In this paper, we propose to deploy a short video recommendation framework on mobile devices to solve these problems. Specifically, we design and deploy a tiny on-device ranking model to enable real-time re-ranking of server-side recommendation results. We improve its prediction accuracy by exploiting users' real-time feedback of watched videos and client-specific real-time features.

With more accurate predictions, we further consider interactions among candidate videos, and propose a context-aware re-ranking method based on adaptive beam search. The framework has been deployed on Kuaishou, a billion-user scale short video application, and improved effective view, like and follow by 1.28\%, 8.22\% and 13.6\% respectively.
\end{abstract}


\maketitle

\section{Introduction}

Short video applications like TikTok, YouTube Shorts, and Kuaishou have grown rapidly in recent years. They have attracted billions of users to create, share and enjoy videos in their daily lives, and have fulfilled their various needs, such as entertainment, learning, or simply killing time, with massive and diverse content. 

\begin{figure}[t]
    \centering
        \begin{subfigure}[b]{0.43\linewidth}
        \centering
    \includegraphics[width=\linewidth]{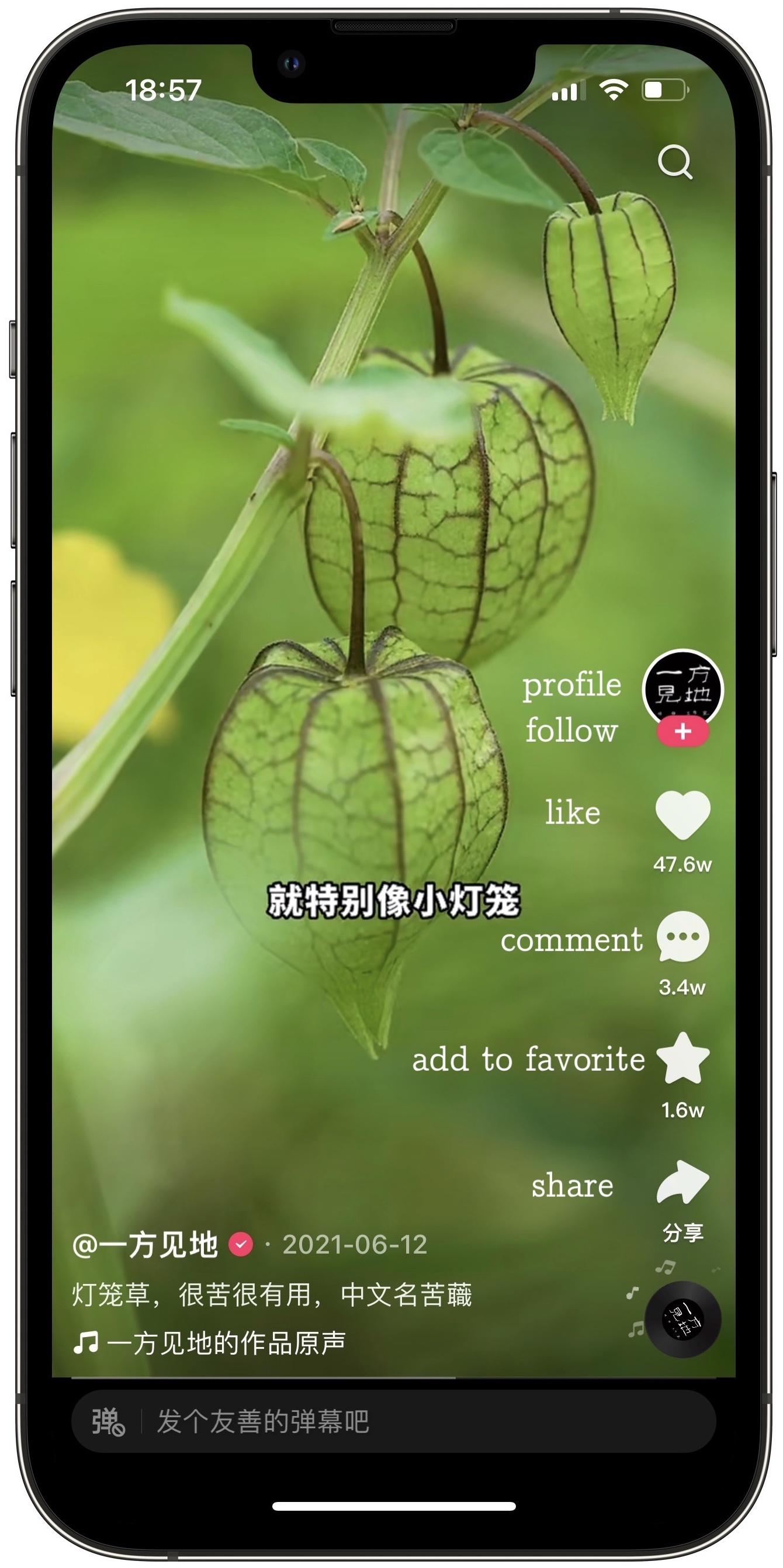} \caption{}\label{fig:kuaishou}
    \end{subfigure}%
    \hfill
    \begin{subfigure}[b]{0.46\linewidth}
    \centering
    \includegraphics[width=\linewidth]{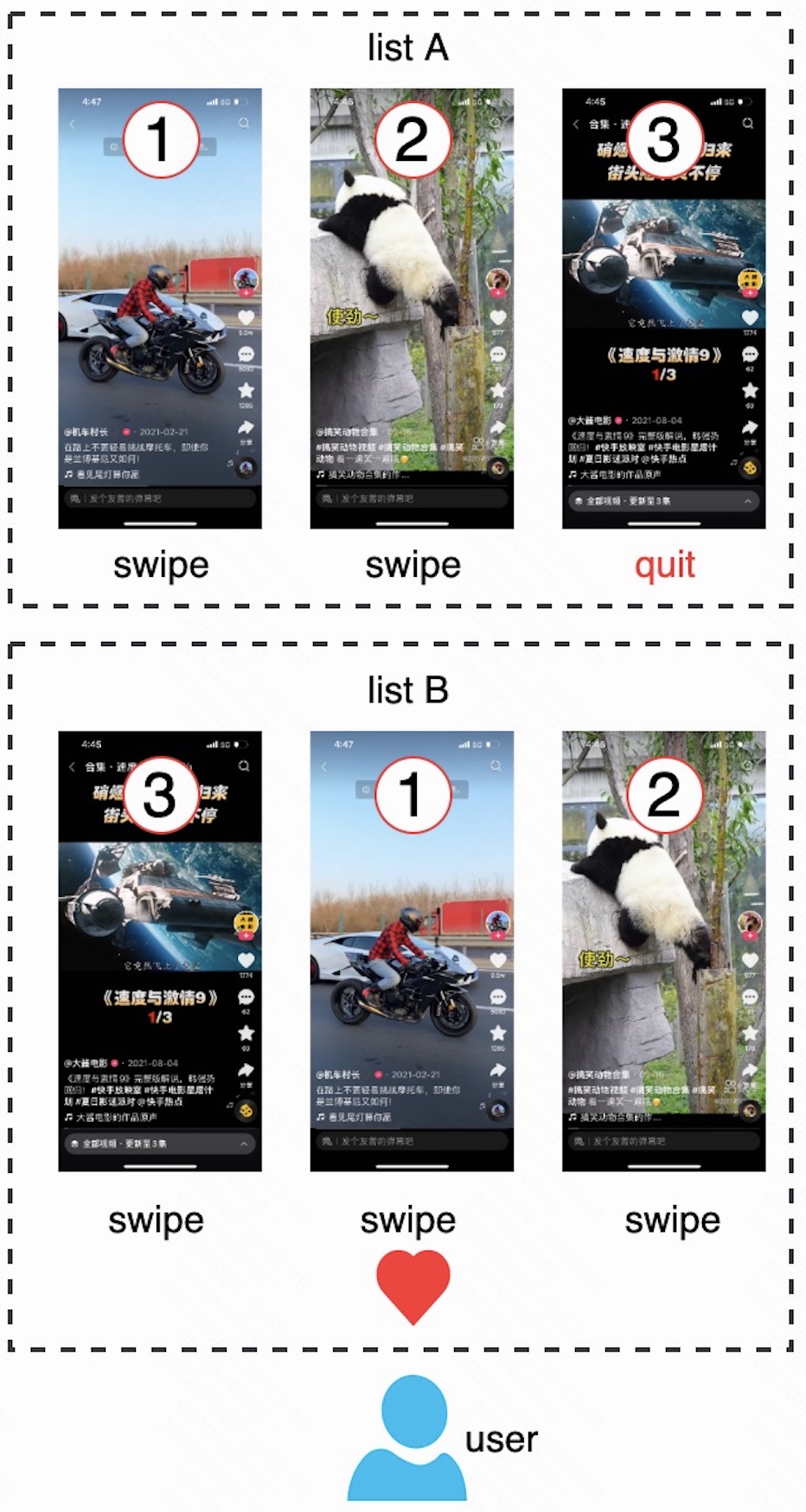}
    \caption{}\label{fig:context}
    \end{subfigure}%
    \caption{\textbf{(a)} Product interface of a short video application example. \textbf{(b)} Example of mutual interactions between videos. Different ordering of the same set of videos will result in different user preferences.}
\end{figure}

In \Cref{fig:kuaishou} is the product interface of a typical short video application, which plays video in full screen mode to create an immersive and distraction-free experience. The interaction between the user and the application is also kept as simple as possible to reduce operation costs. For example, the user can swipe up to switch to the next video, or easily give feedback to videos (e.g., like, comment, add to favorite list, or share with friends) with a simple tap.

Since videos in these applications are short (typically ranging from several seconds to minutes) and diverse, users usually watch a lot of videos on different topics in a short period of time, so that their real-time interests are constantly changing and difficult to predict accurately. As a result, in short video applications, it is very important for the recommender system to be both more accurate and more sensitive to user's real-time feedback.

Traditionally, recommender systems are deployed at the server side, and are generally consisted of multiple stages, such as retrieval, ranking, and re-ranking etc. Since it is such a complicated system, the client usually sends pagination requests to the recommender system to fetch a page of results at once, and display them one by one to the user, in the order decided by the recommender system. After the user finishes watching one page of videos, the client sends another request to fetch the next page, and so on.

There are two main problems in this architecture:
\begin{itemize}
   \item Due to the pagination request mechanism, the recommender system can only interact with client when a new request is sent to the server.
    It is impossible to adjust the content order according to the real-time feedback, even if there may exist some videos that match the user's current interest on the client side.
    \item The real-time feedback from users cannot be exploited immediately. All of the users' feedback must be transmitted to servers before they are usable. 
    Depending on the architecture, the whole process will cost tens of seconds to several minutes, which will hurt the timeliness of the collected feedback data.
    There are also some client-specific features (e.g., the position where the candidate will be displayed, user's current network condition etc.) that are not available in the cloud. All these features are important for real-time context perception and user behavior prediction.
\end{itemize}

With the rapid increase of computational power and storage capability on mobile devices such as phones and tablets, as well as the development
of mobile deep learning frameworks (e.g. TFLite and CoreML), it is possible to offload part of DNN model inference and even training on these devices \cite{gu_server-based_2021,dhar_survey_2021,cai_enable_2022}. A natural benefit is that some lightweight models can be deployed on mobile devices, to provide real-time ranking capability, thus solve the above two problems. It can react immediately to users' implicit (such as watching a video longer than a threshold) or explicit (such as liking or sharing a video) feedback, to make adjustment to remaining candidates accordingly. It is also able to make use of real-time features and client-specific features without any latency, to keep track of the changing context and improve model prediction accuracy.

In this paper, we aim to articulate the design philosophy and architecture choices of a mobile recommender system specifically targeted at short video recommendation scenario.
We emphasize the following lessons learned along the way to successfully deploy such system in a billion-user scale short video application.

\textbf{Feature engineering on real-time signals}. As mentioned previously, the key advantage of client-side recommendation is that we can utilize users' real-time behaviors and some other signals that are not available at the server. This coincides with our empirical results: the apparently appealing idea of edge-cloud 
collaborated model does not lead to significant improvements in our scenario while incurring an inevitable amount of computation and communication overheads; whereas, we find that, by feeding those real-time and complementary signals only along with server-side predictions (such as predicted rates of effective views, likes, follows etc.) into a very lightweight edge-side model, user engagement metrics get substantially improved. Inspired by this observation, we conducted extensive feature engineering to achieve the full potentials of these complementary signals and presents the most effective features and techniques (such as constructing fine-grained crossing features from user feedback) 
adopted in our system (\autoref{sec:features} and \autoref{sect: feature engineering}).

\textbf{Real-time triggered context-aware re-ranking}. The greedy point-wise ranking is not aware of the mutual interactions among recommended videos so that it is only locally optimal. As shown in \Cref{fig:context}, different ordering of the same set of candidates will result in different user preferences. 
To get better ranking result, we need to consider not only immediate reward of the current candidate video, but also its influence on subsequent videos.
List-wise re-ranking approaches provide promising solutions to search for the best possible permutation of candidates with optimal total reward. However, deploying these approaches on the server side suffers from delayed and incomplete contextual information along with high time delay.
On mobile devices, users usually can only see a very limited number $n$ of videos at a time ($n=1$ in our immersive scenario as in \autoref{fig:kuaishou}), so the edge-side re-ranking only needs to determine the next $n$ videos.
Once the user finishes watching these videos, another re-ranking process can be triggered to order the following $n$ videos. This setting brings about better opportunities for us to consider the mutual interactions among videos. We approach this problem by finding a partially ordered list that approximates the optimal one, and propose an efficient technique for context-aware re-ranking, which uses novel adaptive beam search to reduce the searching complexity (\autoref{sec:list_generation}).

Our contributions can be summarized as follows:
\begin{itemize}
    \item We present an edge-side re-ranking solution for short-video recommendation to leverage valuable real-time signals only available on mobile devices and overcome intrinsic limitations of traditional server-side recommender systems. 
    \item We share unique and important lessons (design philosophy, architecture choices, model designs, feature engineering, etc.) we learn when deploying the presented solution in a billion-user scale short-video recommendation platform with non-negligible practical constraints.
    \item We argue that the presented edge-side solution allows a better opportunity for context-aware re-ranking. To take full advantage of it, we propose a novel context-aware re-ranking algorithm specifically tailored for edge scenarios.
    \item We conducted extensive experiments and provide insightful experimental analysis on a real-world industrial scenario. Both offline and online results demonstrate the effectiveness of our edge-side re-ranking solution.
\end{itemize}

\section{Related Work}
There are two lines of existing work that are related to ours: ranking methods and recommender systems on mobile devices.

\subsection{Ranking in Recommendation}
Most of the proposed ranking methods can be classified into three categories: point-wise, pair-wise or list-wise.

\textbf{Point-wise ranking} \cite{cheng_wide_2016,convigton_deep_2016} generally models the ranking problem as a regression (e.g., predict user's rating of a video) or classification (e.g., predict whether the user will like a video) task. The model only use features from the predicting item (aside from common features such as user side features), and no features from other candidate items are used. Point-wise ranking is the most widely used ranking method in recommender systems, however it ignores mutual influences of the candidate items.

\textbf{Pair-wise ranking} \cite{burges_rank_2005,koppel2019pairwise} uses pair-wise loss functions to learn the semantic distance of a pair of items, thus incorporating mutual information from candidate item pairs into modeling. Pair-wise ranking still ignores the contextual information of the whole list, thus is sub-optimal.

\textbf{List-wise ranking} can be further divided into two categories: \textbf{a)} directly optimize list-wise evaluation metrics, such as LambdaMART \cite{burges2010ranknet} to optimize NDCG; or \textbf{b)} consider the mutual influence of items in the input candidate set (either ordered or unordered) to learn a list-wise contextual representation for more accurate prediction \cite{pei_personalized_2019,ai_learning_2018,feng_grn_2021}. We mainly focus on the latter category in our work. Compared with point-wise and pair-wise ranking, list-wise ranking has the advantage of capturing more accurate contextual information to improve model performance. However, it also has the highest computational complexity, and needs to be simplified to meet latency requirements in production.

In our scenario, we adopt a novel context-aware planning method based on adaptive beam search for re-ranking.

\subsection{Recommendation on Mobile Devices}
With the increasing computational power of mobile devices, some practitioners have been researching the possibility of deploying ranking models directly on the client side, for better utilization of real-time features, such as user feedback. EdgeRec \cite{gong_edgerec_2020} is the first attempt to deploy ranking model on mobile devices to reduce signal latency, and achieves obvious gain, demonstrating the effectiveness of real-time features. Although EdgeRec has a context-aware re-ranking module, which uses a GRU structure to encode all the candidate items in the initial order to get a local ranking context, it ignores that if the item order is changed by re-ranking, the local ranking context will also change, thus it is not accurate anymore. Besides, in our system, the page size is typically less than 10, and is far smaller than that in EdgeRec (which is 50), which means our server-side recommendation systems have more chances to interact with the client to achieve faster user interest adaption. A lot of engineering effort has also been made to reduce data transmitting latency in the whole system. Thus it is more challenging to improve on an already very high base.

\cite{yao_device-cloud_2021} proposes a device-cloud collaborative learning framework, which learn a patch model on device to achieve personalized ranking model, and update the centralized cloud model by aggregating patch models from different devices. \cite{gu_-device_2022} also aims to improve model personalization by combining local data set of each user and similar samples retrieved from cloud to train ranking model on device. \cite{muhammad_federated_2019,muhammad_fedfast_2020} use Federated Learning \cite{mcmahan_communication_2017} to train recommendation models collaboratively for better privacy protection, because training data is only locally accessed at each client without being transferred to the server.

Different from the above work, our focus is on designing a tiny model that fits on mobile devices, and we pay special attention to the use of real-time features by designing dedicated feature engineering techniques.

\section{System Overview}
\begin{figure}
  \includegraphics[width=\linewidth]{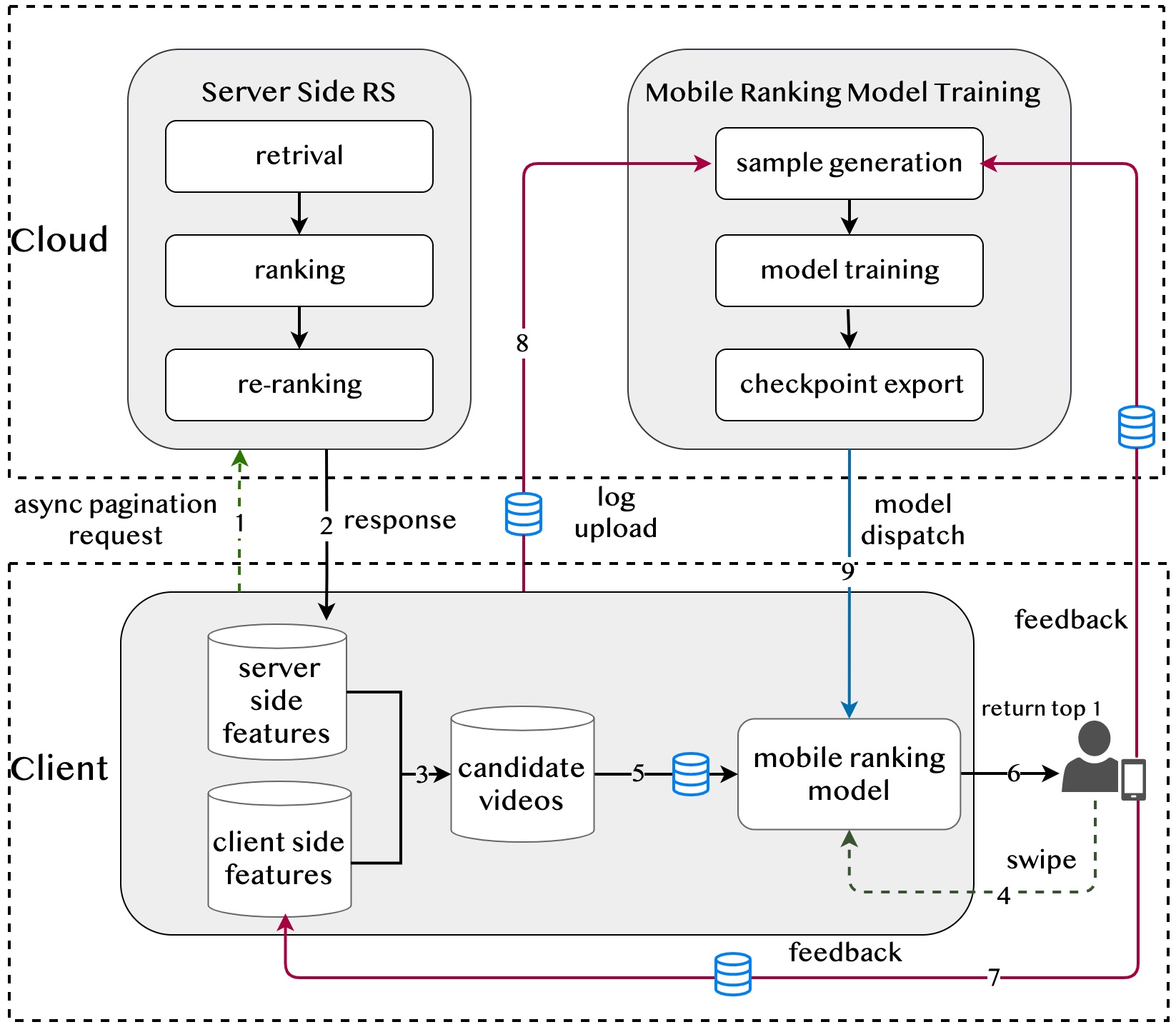}
  \caption{Architecture of proposed short video recommender system on mobile devices.}\label{fig:system_architecture}
\end{figure}

The whole framework can be divided into three modules, as show in \autoref{fig:system_architecture}.

\subsection{Server-Side Recommendation System}
The first module is a traditional recommender system deployed on the server side. It is consisted of retrieval, ranking, and re-ranking stages. The result size of each stage is generally on the order of thousands, hundreds and tens, respectively. When client initiates a pagination request, server will go through these stages to generate an ordered list of recommended videos. Some server-side item features of the recommended candidates (a subset of the input features to the ranking model on mobile devices), such as predicted scores from server-side ranking model, will be extracted to send along with these candidates to client. In our system, prior to deploying ranking model on mobile devices, server will send $m$ candidate videos in response to client, and when these videos are consumed by user, a new request will be sent to the server to fetch another $m$ videos. With client-side ranking model, the server will send extra $n$ videos to increase candidate space. The client still sends a new request once every $m$ videos are consumed, and the rest $n$ videos not shown are discarded.

\subsection{Model Training System}
The second module is a model training system. Similar to other such systems, it first generates training samples from collected data; then
use distributed training to train the ranking model in an incremental way. The checkpoint is exported periodically, and converted to TFLite format for deployment.
The details of the client-side ranking model will be introduced in \autoref{sec:model_architecture}.
\subsection{Client-Side Recommendation System}
The third module is a recommendation system deployed on the client-side. It can be further divided into two parts:

\textbf{Feature collection}. This part collects features from both server side and client side, then joins them together to form complete input feature set to be sent to ranking model. Specifically, the client maintains a watched video list, and all the features and user feedback of each video in the list will be collected and stored. Every time a video is consumed, it will be appended to the list, so we can extract real-time signals from this list with almost no latency.

\textbf{Context-aware re-ranking}. When the user swipes to watch next video, or likes/shares a video, the system will trigger the model on device to re-rank the candidates according to the user's behavior. These triggers are configurable in our system, and in production, only swipe is currently used to trigger re-rank. When the re-rank process is triggered, the client first generates input features from both watched video list and candidate set, then feed the input to re-ranking model, and a context-aware ranking method is used to sequentially generate an ordered list with largest ListReward as defined in \autoref{eq:listreward}. After re-ranking, client inserts the top-ranked video at the next position.

This module also uploads logged data to the server-side for model training and data analysis.

\section{On-Device Ranking Model}\label{sec:model_architecture}

\subsection{Design Philosophy}
Since this model is deployed on mobile devices, due to the storage, computational power and energy consumption constraints, it has to be extremely lightweight yet effective. When designing the model architecture, there are mainly two choices: \textbf{a)} a large edge-cloud collaborated model that keeps embedding parameters (which generally comprises most part of the parameters in the model) on the server side, and only send parameters of DNN layers to client. When doing inference, the server first looks up needed embeddings, and send them to client for following computation; or \textbf{b)} a carefully designed small model that fits in mobile devices in its entirety. We choose the second way, i.e., design a small but self-contained model for mobile devices. This model is a complement of server-side model, in the sense that it mainly takes advantage of client-side user real-time feedback to improve prediction accuracy. Ranking model on server side has compressed most of the information into the final prediction scores, so we can use this as input to avoid redundant computation, and make the model small enough. This not only reduces computation latency, but also get rid of the need to keep multiple versions of model to ensure consistency between client and server in the split-model setting. Our design decision is also supported by offline experiment, where a large model with complicate features (such as video id and user id) and model structure does not bring obvious improvement compared with a small one. We conjecture the reason is that these features have already been used in server-side ranking model, thus there is little extra information in the input.

\subsection{Input Features}\label{sec:features}
In this subsection, we will introduce the input features to the model, and some feature engineering techniques specifically designed for real-time features are followed in the next subsection.

Because we choose to design a tiny model that fits in mobile devices, we have to carefully choose the most important features to keep the model as small as possible.

These features can be classified into 3 categories:

\textbf{Server-side prediction}. This is one of the most important features used in edge ranking model. Server-side ranking model is complicated in the sense of both feature system and model structure. Especially in our system, this model is trained in an online learning fashion using streaming data, and it uses a lot of ID features (such as video id, user id, and crossing features), and users' watch history in a fairly long time, so it is good at capturing user's long term interest. Its predictions contain highly condensed information distilled from input features, such as whether the user will like current video. We can use client-side real-time features as a complement to better perceive user's real-time interest.

\textbf{Video static attributes}. Every video has many static attributes, such as video id, category, duration, tag, description, cover image, background music, etc. All of them are very important to help model learn, however due to the size limit, we can only use a small subset of them. In our experiment, we only use video category and duration attributes, which have less than 10,000 distinct values combined (video duration is cut off at 1800 seconds), so it will not increase the model size much.

\textbf{Client-side features}. During running, client will collect many important features, such as user feedback, video watch time etc. These features are attached with corresponding video, and stored in watched video list with a limited length, since we focus on real-time features, and previous, older user feedback are already used in server-side ranking model. There is also one specific set of features that are only accessible on edge, such as the position where the candidate will be displayed, or current network condition. As another example, users will watch videos under different network conditions (e.g., on public transport such as subways, where signal is unstable), and to avoid intermittent playback caused by network instability, client will buffer a small part of each candidate video in advance, and the buffer length is also an important feature. Traditionally, they are treated as bias features \cite{zhuang_globally_2018,zhao_recommending_2019} (or ``privileged feature'' in \cite{xu_privileged_2020}) which are only used when training ranking models, and are treated as missing at serving time, because we cannot know their value on the server side (\cite{xu_privileged_2020} tries to distill the information of such features into a student model, however there is still an obvious gap in performance). Instead, when we do inference on mobile devices, they become readily available, and carry important information. We can see in the experiment that they indeed help to greatly improve model performance.

We summarized features used in our model in \autoref{tab:feature}.

\begin{table}
    \caption{Features used in mobile ranking model.}
    \label{tab:feature}
    \begin{tabular}{ccc}
    \toprule
    \textbf{feature} & \textbf{source} & \textbf{description} \\ \midrule
     & & various rates predicted by server-side \\
    $pXTR$ & server & ranking model, such as $pLTR$ for \\
    && predicted rate of user liking a video. \\ \midrule
    $v_d$ & video & video duration \\ 
    $v_c$ &  & video category \\ \midrule
    $v_w$  &  & watch time of watched videos \\ 
    $v_{feedback}$ & & user feedback such as like, share etc. \\
    $v_t$ &  &video impression timestamp \\ 
    $v_{pos}$ & client &video impression position \\
    $u_{net}$ &  & current net condition \\
    $v_{buffer}$ &  & buffered length of candidate videos \\
    \bottomrule
    \end{tabular}
\end{table}
\renewcommand{\thefootnote}{\fnsymbol{footnote}}
\subsection{Feature Engineering}
\label{sect: feature engineering}
To enhance the influence of real-time features, we add crossing features derived from them in model inputs. Specifically, we add the following crossing features:
\begin{itemize}
    \item $pXTR$ diff, which is calculated as $pXTR - pXTR^h$\footnote[2]{Superscript $h$ indicates the corresponding feature of an item in history list.}. The intuition is that, with $pXTR$ diff as input feature, the model can perceive user's preference shift in real-time. For example, if in current session, the user does not give positive feedback to several videos in a category with high $pXTR$, then maybe she is not interested in videos from such category currently. So for the rest candidates in the same category, if their $pXTR$ is relatively low (i.e., with a negative $pXTR$ diff score), they should not be ranked to the top. On the other hand, if the user likes videos in a category with relatively low $pXTR$, we can try to increase the probability of showing videos from the same category with higher $pXTR$ (i.e., with a positive $pXTR$ diff score), because the user probably enjoys such type of videos at the moment. Instead of using a static score as anchor (such as average $pXTR$ of user engaged videos in the past), using $pXTR$ of recently watched videos can automatically adapt to user's real-time interests.
    \item Time since last impression, which is calculated as $v_t - v_t^h$. This is to capture the temporal importance of previously watched videos. Generally, the more recent an impression is, the more influential it will be.
    \item Impression position gap between videos, which is calculates as $v_{pos} - v_{pos}^h$. This is similar to temporal diff, but it only considers impression position, which will be more stable if the user consumes videos at varying speed.
\end{itemize}

They are all further crossed with video category and user feedback to capture user's fine-grained preferences.

Their effectiveness will be reported in \autoref{sec:exp}.

\subsection{Model Architecture}

\begin{figure}
  \includegraphics[width=\linewidth]{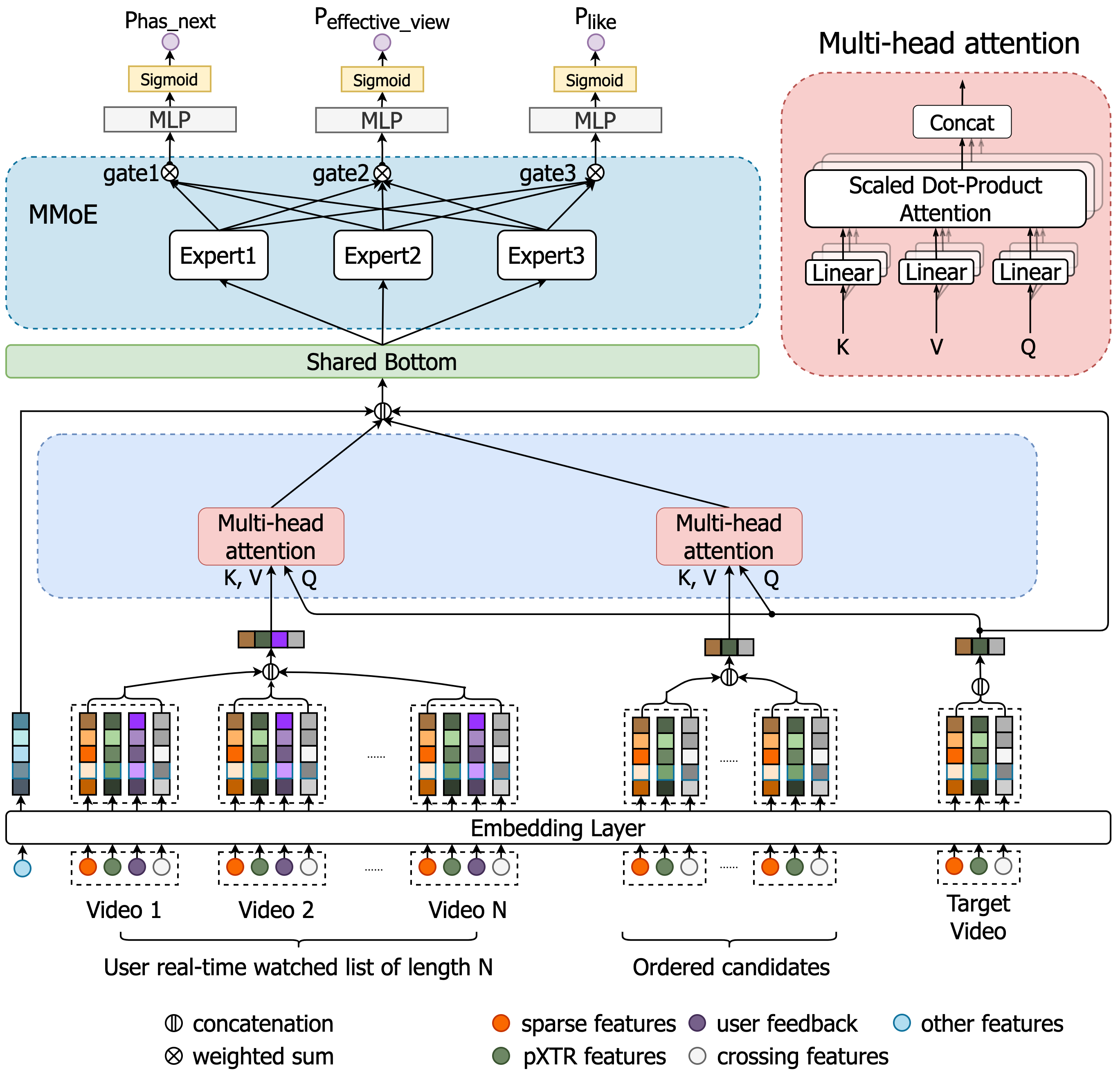}
  \caption{Architecture of the on-device ranking model.}\label{fig:model}
\end{figure}

In recommendation, mutual influence among items will lead to different user preferences for different ordering of the same set of candidates, which has been shown in previous work \cite{zhuang_globally_2018,pei_personalized_2019,ai_learning_2018,gong_edgerec_2020,feng_grn_2021}. So it is important for the model to incorporate such mutual influence to be aware of the ranking context. Here we define the ranking context of video $v_i$ in an ordered candidate list $\mathbb{L}$ as
\begin{equation}\label{eq:context}
    c(i) = c(\mathcal{H};v_1,v_2,\dots,v_{i-1};\mathcal{O}),
\end{equation} where $\mathcal{H}$ is the watch history sequence. $v_1,v_2,\dots,v_{i-1}$ are ordered candidates before $v_i$. $\mathcal{O}$ represents other contextual information, such as the user's current net condition etc. This means that user preference for $v_i$ is influenced by many different factors, and all of them should be considered by the ranking model.

Considering this, the architecture of our mobile ranking model is presented in \autoref{fig:model}, which has 4 types of inputs: 
\begin{itemize}
    \item \textbf{Real-time watch history sequence} is the client-side maintained real-time watched video list. It includes both video information and corresponding user feedback.
    \item \textbf{Ordered candidates list} is added to help model the interactions between ordered candidates and target video, in order to facilitate the context-aware planning method introduced in \autoref{sec:list_generation}. When training, ordered candidates are chronologically ordered previously watched videos of current user, with max length in accordance with max beam search steps.
    \item \textbf{Target video} is the video to be predicted.
    \item \textbf{Other features} are mainly contextual features such as impression position and network condition etc.
\end{itemize}

The watch history sequence is modeled using a multi-head attention (MHA) \cite{vaswani_attention_2017} module with target attention \cite{zhou_deep_2018}, calculated as follows:
\begin{equation}
    \mbox{Attention}(\boldsymbol{Q},\boldsymbol{K},\boldsymbol{V}) = \text{softmax} \left( \frac{\boldsymbol{Q}\boldsymbol{K}^T}{\sqrt{d}} \right) \boldsymbol{V}
\end{equation}

$\boldsymbol{Q}, \boldsymbol{K}, \boldsymbol{V}$ are the query, key and value, respectively. $d$ is the embedding dimension. The query $\boldsymbol{Q}$ is projected from features of the candidate item, while key $\boldsymbol{K}$ and value $\boldsymbol{V}$ are both projected from features of watch history sequence.

To explicitly model influence of already ordered candidates on current target video to be predicted, we use another MHA module with target attention, in which the key $\boldsymbol{K}$ and value $\boldsymbol{V}$ are projected from features of ordered candidates. Since we want to model the \textit{immediate} reward of target video, we do not add remaining candidates in model inputs, because they can hardly affect the result \cite{zhuang_globally_2018}. This is verified in our offline experiments, and for brevity the result is not presented here.

The outputs of two MHA modules are concatenated with other features and target video features to form input to a Multi-gate Mixture-of-Experts (MMoE) \cite{ma_modeling_2018} module. Each task uses a different gate to combine expert outputs to go through a feed-forward network and a sigmoid function to get the final prediction.
\subsection{Model Learning}
In our scenario, there are many targets to consider, including watch time, user interactions (e.g., like, share, comment), etc. Since multi-task learning is not our focus in this paper, we omit the details for brevity, and choose 3 binary targets closely related to the quest of improving users' satisfaction to our platform as learning goals. These 3 targets are ``has\_next'', ``effective\_view'', and ``like''. ``has\_next'' is defined as the user continues to watch videos after current one; in immersive scenario where video will automatically start playing in full screen mode, there is no ``click'' operation, so we define an ``effective\_view'' label as user watches a video longer than a threshold (e.g., 5 seconds), and videos in different duration intervals have different thresholds; ``like'' is defined as the user likes current video by clicking the like button or double tapping/long pressing screen.

We train the model in a multi-task learning fashion. The loss function is defined as the sum of log losses of each target, averaged by the number of training samples:
\begin{equation}
    \mathcal{L}(\Theta) = -\frac{1}{N}\sum_{i=1}^{N} \sum_{j=1}^3 w_j \left (y_{ij}\log \hat{y}_{ij} - (1-y_{ij})\log(1-\hat{y}_{ij}) \right ),
\end{equation}
where $\Theta$ is the set of model parameters, $N$ is the total number of training instances, $w_j$ is the weight of loss $j$ and is manually tuned and kept the same in all the experiments, $y_{ij} \in \{0, 1\}$ is the $j$-th ground-truth label of item $i$, $\hat{y}_{ij}$ is the $j$-th predict the result of item $i$. We optimize $\Theta$ by minimizing $\mathcal{L}(\Theta)$ through gradient decent.

\subsection{Deployment}
We export model checkpoints periodically in the training process, then convert the checkpoint to TFLite format, and upload it to CDN, along with its MD5. When a client starts, it will upload the MD5 of the local model file to the server, and the server compares it with the MD5 of the current model. If the client-side model is outdated, then the server tells the client to download the new model.

\section{Real-time Triggered Context-Aware Re-ranking}\label{sec:list_generation}
Once a user finishes watching a video and generates new real-time ranking signals, we can responsively update our client-side model predictions and trigger a new re-ranking process. 
Given the updated model predictions, there are many ways to determine the list of videos to show users. The most widely used is point-wise ranking, which greedily orders the videos by their scores decreasingly. However, point-wise ranking ignores the mutual influence among candidates, thus is not optimal. 

Ideally, we want to find the optimal permutation $\mathcal{P}$ of the candidate set $\mathcal{C}$, which leads to maximum ListReward (LR) defined as:
\begin{equation}\label{eq:listreward}
    \mbox{LR}(\mathcal{P}) = \sum_{i = 1}^{|\mathcal{P}|} s_i \left ( \alpha p(\mbox{effective\_view}_i \mid c(i)) + \beta p(\mbox{like}_i \mid c(i)) \right),
\end{equation}
where
\begin{equation} \label{eq:next}
    s_i = \left \{ 
    \begin{array}{ll}
    \prod_{j=1}^{i-1} p(\mbox{has\_next}_j \mid c(j)),& i \geq 2 \\
    1,& i=1 \\
    \end{array}
    \right.
\end{equation} is the accumulated has\_next probability till position $i$, which acts as a discounting factor to incorporate future reward. $p(\mbox{has\_next}_i \mid c(i))$, $p(\mbox{effective\_view}_i \mid c(i))$, and $p(\mbox{like}_i \mid c(i))$ are the predictions on has\_next, effective\_view, and like of $v_i$ respectively, considering ranking context $c(i)$ defined in \autoref{eq:context}. $\alpha$ and  $\beta$ are weights of different rewards. 

However, directly searching for the optimal permutation requires evaluating the reward of every possible list, which is prohibitively expensive, since it is of factorial complexity $O(m!)$ ($m$ is the size of candidate set). Beam search is a commonly used approximation solution to such problem, which reduces the time complexity to $O(km^2)$, where $k$ is the beam size. Yet, the quadratic time complexity is still too high to deploy in our production environment. Fortunately, different to the case of server-side re-ranking, we only need to lazily determine the next $n$ videos users can simultaneously see on their devices ($n=1$ in our immersive scenario as shown in \autoref{fig:kuaishou}). Also, in our offline experiment (\autoref{tab:beamsearch}), we observe that the relative difference of ListReward among different beams in each search step decreases monotonously as the search step grows. Thus, we propose a novel beam search strategy to choose an adaptive search step $n\leq l \ll m$ and further reduce the searching time complexity to $O(klm)$.

\begin{figure}
    \centering
    \includegraphics[width=\linewidth]{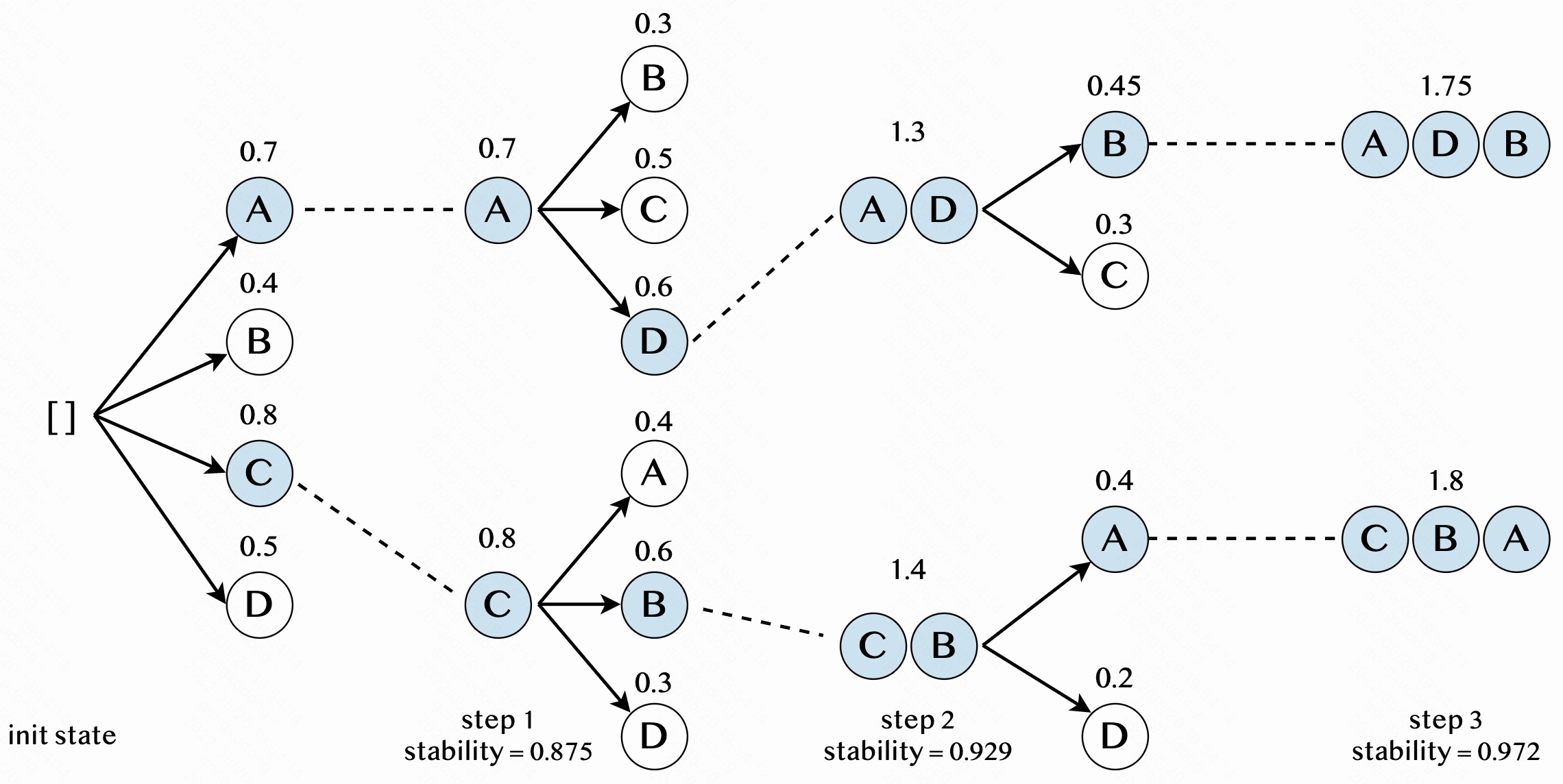}
    \caption{Illustration of the adaptive beam search process with number of candidates $n = 4$, beam size $k = 2$, and stability threshold $t = 0.95$. The number above each candidate or candidate list is the corresponding reward.} \label{fig:beam_search}
\end{figure}

To realize adaptive beam search, we define a \textit{stability} measure as the minimum ListReward divided by the maximum ListReward in the current beam search step.
\begin{equation}\label{eq:stability}
\mbox{stability}(score\_list) = \frac{\min(score\_list)}{\max(score\_list)}
\end{equation}
Once the stability exceeds a given threshold $t$, the beam search process is terminated to save unnecessary computation, since we can expect there will not be a large difference in the remaining search steps. The adaptive beam search process is illustrated in \autoref{fig:beam_search}.

The algorithm is sketched in \autoref{algo:list_generation}, and it is implemented inside the exported TFLite execution graph.
\begin{algorithm}
  \caption{Context-aware re-ranking with adaptive beam search}\label{algo:list_generation}
  \begin{algorithmic}[1]
    \Require candidate videos set $\mathcal{C} = \{v_1, v_2, \dots, v_l \}$, ranking model $\mathcal{M}$, beam size $k$, number of videos $n$ to show next simultaneously, stopping stability threshold $t$
    \Ensure Next video to show to the user
    \State beam\_indices $\gets$ [[$-1$] for \_ in range($k$)] \Comment{initialize}
    \State beam\_scores $\gets$ [[$-\infty$] for \_ in range($k$)]
    \For {$i \gets 1$ to $l$}
        \State features $\gets$ \textsc{GenerateFeatures}($\mathcal{C}$, beam\_indices)
        \State beam\_predicts $\gets$ $\mathcal{M}$(features)
        \State list\_scores $\gets$ \textsc{ListReward}(beam\_predicts) \Comment{\autoref{eq:listreward}}
        \State beam\_indices, beam\_scores $\gets$ \textsc{TopK}(list\_scores, $k$)
        \If {$i \ge n$ \&\& stability(beam\_scores) $\geq t$} \Comment{\autoref{eq:stability}}
            \State break
        \EndIf
    \EndFor
    \State $\mathbb{L} \gets$ top indices list in beam\_indices
    \State \Return first $n$ elements in $\mathbb{L}$
  \end{algorithmic}
\end{algorithm}

\section{Experiments}\label{sec:exp}
In this section, we conduct experiments to evaluate the offline and online performance of both the ranking model and the recommendation method.

For the ranking model, we want to evaluate the effect of client-side real-time features, including user feedback and other features not available on the server side. For the recommendation method, we compare our proposed context-aware re-ranking method with greedy point-wise re-ranking in the online environment to show the effect of ranking context.

\subsection{Offline Experiment}
\subsubsection{Dataset}
We collected data from a large short video recommendation system in 8 consecutive days for offline evaluation. The first 7 days are used for training, and the last day is used for test. A summary of the dataset is shown in \autoref{table:dataset}.
\begin{table}
  \caption{Dataset summary}\label{table:dataset}
  \begin{tabular}{c|ccc}
  \toprule
  \textbf{Dataset} & \textbf{\#Users} & \textbf{\#Videos} & \textbf{\#Records} \\ \midrule
  Train & 2,421,196 & 4,030,717 & 60,142,557 \\
  Test & 1,098,351 & 1,614,608 & 18,538,868 \\
  \bottomrule
  \end{tabular}
\end{table}

\subsubsection{Evaluation Metrics and Baselines}

For the ranking model, we use AUC of each target as evaluation metric, and compare with following baseline models:
\begin{itemize}
    \item \textbf{ServerScore}, which is calculated using the logged prediction scores from the server-side ranking model.
    \item \textbf{SimpleDNN}, which is a simple DNN Model with all the real-time features, except for feature engineering techniques proposed in this paper. 
    \item \textbf{EdgeRec}, which is the model proposed in \cite{gong_edgerec_2020}. We implement it as described in the paper, with all the real-time features, except for feature engineering techniques proposed in this paper. 
\end{itemize}

Since there are 3 targets in our scenario, and EdgeRec is a single-task model in original setting, it is modified by replacing the final MLP and outputs layers with an MMoE module and 3 towers, one for each task. The configurations of hidden layer sizes are the same as these in our model.

\subsubsection{Experiment Setup}
Features of float type are discretized and embedded using the AutoDis  \cite{guo_autodis_2021} technique, with embedding size 8. We find this gives slightly better results in our experiment. User feedback are embedded to 8-dimensional vector, while duration and category are embedded to 16-dimensional vector.

The MHA module has 8 heads, and dimension of each head is 16. The MMoE module has 12 experts with hidden size 64. The tower of each task is a four-layer MLP with hidden size $[128, 64, 32, 1]$. We use ReLU as activation function, and all the models are randomly initialized, and trained in an end-to-end manner, using Adam optimizer \cite{diederik_adam_2015} with batch size 1024 and learning rate 0.001.

\subsubsection{Offline Result}
\begin{table}[h]
    \caption{AUC of different models.}
    \label{tab:exp}
    \begin{tabular}{c|ccc}
    \toprule
     \multirow{2}{*}{Model} & \multicolumn{3}{c}{AUC}\\
     & has\_next & effective\_view & like \\
     \midrule
      ServerScore  & - & 0.7728 & 0.9483  \\
      SimpleDNN & 0.709 & 0.766 & 0.9185 \\
      EdgeRec & 0.719 & 0.7812 & 0.9432 \\ 
      Ours & \textbf{0.7293} & \textbf{0.7884} & \textbf{0.9496} \\
    \bottomrule
    \end{tabular}
\end{table}

The experiment results are reported in \autoref{tab:exp}. Because server-side ranking model does not predict has\_next target, the corresponding metric is not reported.

From the result, we can see that the performance of SimpleDNN on effective\_view and like is even worse than the ServerScore, showing the difficulty of utilizing the real-time features within a tiny model. EdgeRec is better than SimpleDNN on all metrics showing the importance of suitable network architecture, though it still has lower AUC on like compared to ServerScore. 
Our model achieves the best performance on all the metrics, which demonstrates the effectiveness of real-time features under proper feature engineering and network design techniques.

\subsubsection{Ablation Study}
Ablation study includes two parts: the effect of real-time features and feature engineering techniques in ranking model, and the effect of search step on beam search stability.

\paragraph{Ranking Model}
We conduct ablation study for different part in the model:
\begin{itemize}
    \item Client-specific features (\textbf{CSF}), which include impression position and buffered length of the target video, and current net condition.
    \item Feature engineering techniques (\textbf{FE}), including various feature crossing and AutoDis.
    \item Real-time sequence of watched videos (\textbf{RTS}), which contains latest watched videos with corresponding feedback.
\end{itemize}
\begin{table}[h]
    \caption{Result of ablation experiment on different parts in ranking model.}
    \label{tab:ablation}
    \begin{tabular}{c|ccc}
    \toprule
     \multirow{2}{*}{Model} & \multicolumn{3}{c}{AUC} \\
     & has\_next & effective\_view & like \\
     \midrule
      Full Model & \textbf{0.7293} & \textbf{0.7884} & \textbf{0.9496} \\
      Full Model - CSF & 0.704 & 0.7864 & 0.9495 \\
      Full Model - FE & 0.7105 & 0.78 & 0.9213 \\
      Full Model - RTS & 0.7221 & 0.7846 & 0.9492 \\
    \bottomrule
    \end{tabular}
\end{table}

The ablation study experiment results for ranking model are shown in \autoref{tab:ablation}. Compare full model and model without client-specific features (Full Model vs. Full Model - CSF), we can see that removing client-specific features causes 0.0253, 0.002, 0.0001 AUC drop on has\_next, effective\_view and like respectively. The large drop on has\_next confirms the empirical evidence that user's exit probability is highly related with browsing depth and network condition. It also has a strong affection on effective\_view, proving client specific features are important in ranking.

If we remove the feature engineering from model (Full Model vs. Full Model - FE), AUC of effective\_view and like drops more than removing client-specific features, which demonstrates that careful feature engineering has a strong influence on model performance.

Missing Real-time sequential features (Full Model vs. Full Model - RTS) also has a negative impact on model performance, and it is further demonstrated by the case study in \autoref{sec:case}.

\paragraph{Beam Search Stability}

The effect of different search steps on stability in beam search is shown in \autoref{tab:beamsearch}. We can see that the stability of the beam search result (see \autoref{eq:stability} for definition) grows rapidly and monotonously with the increase of the search step. When the search step is longer than 3, the stability reaches above 0.99. This shows that latter search steps have diminishing influence on beam quality, and validates our decision to do a partial beam search of limited search step for comparable result and much better efficiency. We set stopping stability threshold $t$ in \autoref{algo:list_generation} to 0.95 in online A/B testing, which means the average search step is less than 3, and it helps to increase computing efficiency.

\begin{table}[t]
    \caption{Stability of different beam search steps under beam size 4.}
    \label{tab:beamsearch}
    \begin{tabular}{ccc}
    \toprule
     Search Step & Stability & Latency (relative) \\ \midrule
     1 & 0.6715 & 1.0 \\
     2 & 0.9421 & 1.85 \\
     3 & 0.9915 & 2.77 \\
     4 & 0.9939 & 3.91 \\
     5 & 0.9954 & 4.89 \\
     \bottomrule
    \end{tabular}
\end{table}

\subsection{Online A/B Testing}
To further evaluate the effectiveness in online scenario, we deploy models in production environment and test the performance during one week, and the result is presented in \autoref{tab:abtest}.

\subsubsection{Experiment Environment}
The total number of parameters of the model deployed in production is 1.32 million, and the size of generated TFLite model is less than 6MB, so it can be fully deployed on mobile devices. The client will send a new pagination request once every 6 videos are consumed, and for each request, the server will return 9 videos to the client to provide more choices.

The baseline is the production recommender system without mobile re-ranking. We conduct experiments of two different configurations, both using the same mobile ranking model: \textbf{a)} greedily re-ranking candidates in a point-wise way, and \textbf{b)} using context-aware adaptive beam search (\autoref{algo:list_generation}) with beam size 4 and stopping stability threshold 0.95 to search for best video which brings largest ListReward.

\subsubsection{Result and Analysis}
\begin{table}
    \caption{Online A/B testing metrics during one week.}
    \label{tab:abtest}
    \begin{tabular}{ccccc}
    \toprule
      Strategy   &  Effective View & Like & Follow \\ \midrule
      Greedy & +0.907\% & +5.956\% & +11.795\% \\
      Context-aware Re-ranking & +1.277\% & +8.218\% & +13.598\% \\
      \bottomrule
    \end{tabular}
\end{table}

During the A/B testing period, our two experiments consistently outperform base group. Mobile ranking model brings 0.907\% improvement on effective view, 5.956\% improvement on like, and 11.795\% improvement on follow. Beam search brings another 0.37\%, 2.262\% and 1.803\% improvement on these metrics. Improvement at this scale is considered significant in our production system, and it shows that users are more satisfied with videos re-ranked by client-side model. This ranking model has been serving the whole traffic in our production environment.

\begin{figure}
    \centering
    \includegraphics[width=\linewidth]{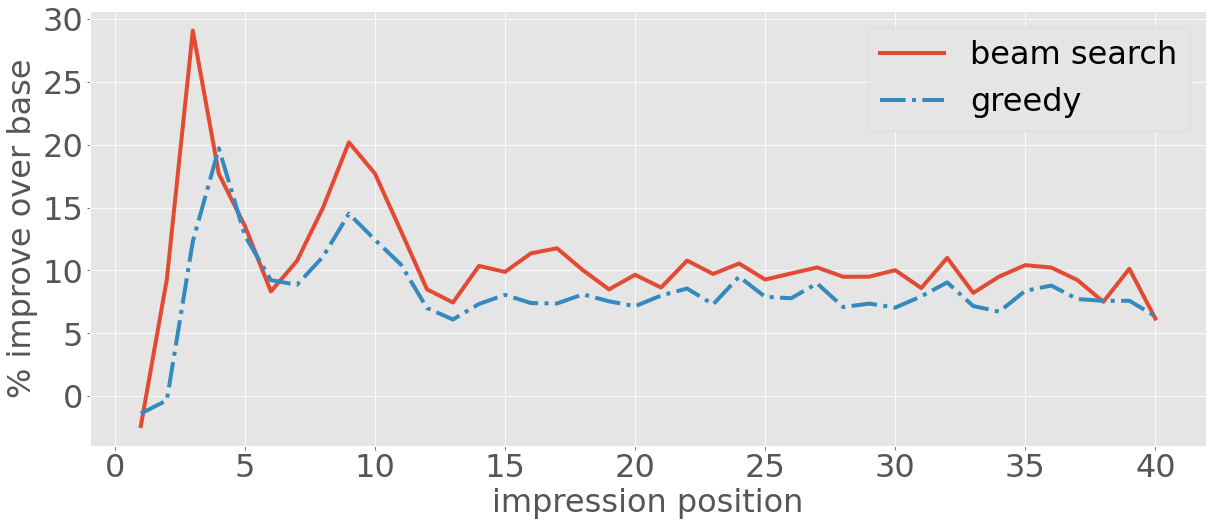}
    \caption{Relative improvement of like over base group in online experiment.} \label{fig:online_evtr}
\end{figure}

We give further analysis of the relationship between relative improvement of like and impression position in each session in \autoref{fig:online_evtr}. Similar phenomenon is observed for other metrics. We can draw following conclusions from the figure. \textbf{a)} At the beginning of each session, the performance of experiment group is the same or even slightly worse than that of base group, because lack of user feedback will affect model prediction. As users watch more videos, performance of both experiment groups increase rapidly, and is consistently higher than base group. This shows that real-time feedback are important for better user perception. \textbf{b)} The improvement is periodic, which first peaks at certain position, then slightly drops. There are two reasons for this fluctuation. First reason is the periodically varying candidate set size. When new candidates arrive, client has more choices at current position, so it can choose a video better satisfying user's real-time interest. As candidates number decreases, the potential gain is also smaller, until next page of candidates are fetched. Second reason is that when server receives new pagination request from client, it can exploit latest received feedback signals to recommend videos better meet current user needs, so at the beginning of each page, advantage of client-side re-ranking is reduced. \textbf{c)} The performance with beam search is consistently higher than greedy re-ranking, which proves the benefit of context-aware planning.

\begin{figure}
    \centering
    \includegraphics[width=\linewidth]{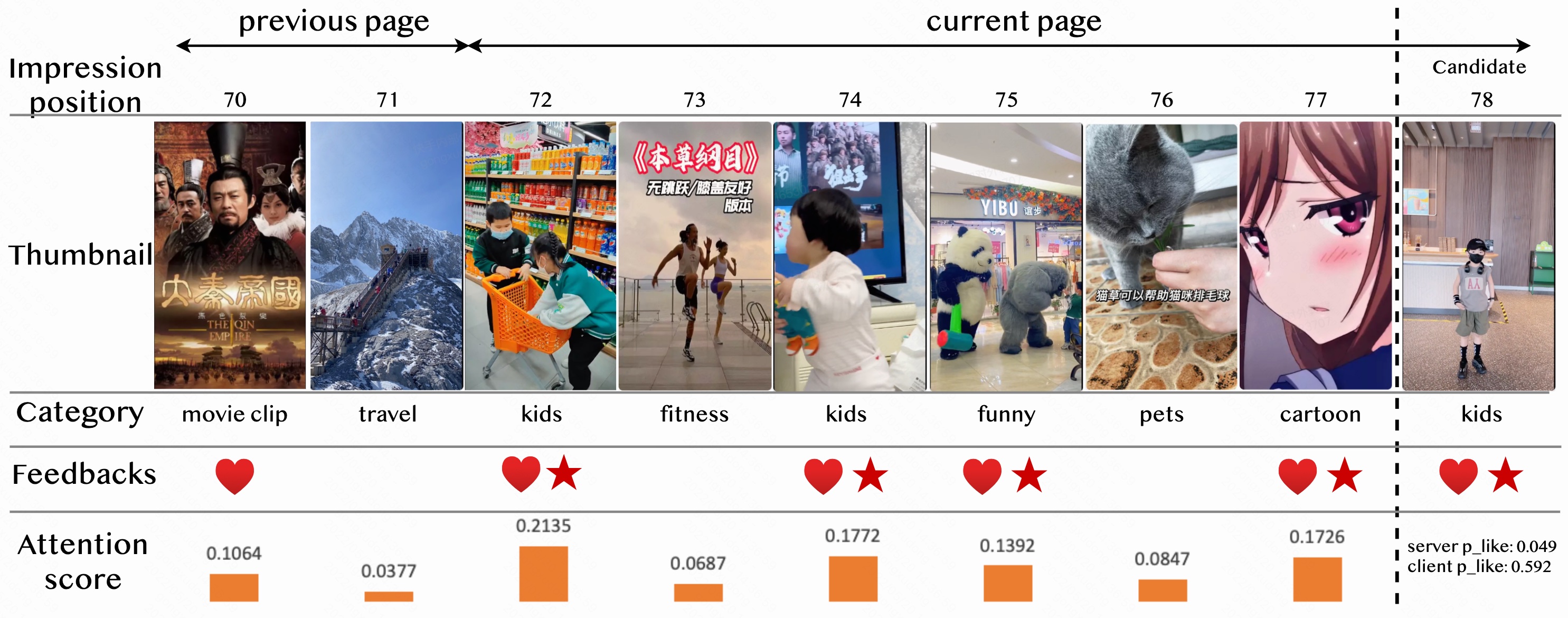} 
    \caption{A case study to show the influence of real-time feedback.} \label{fig:case}
\end{figure}

We also monitored the computing efficiency and resource usage of devices in experiment group, in comparison with base group without ranking model on mobile devices. The result of ranking model with beam search is shown in \autoref{tab:resource}, because it is more complicated and resource-consumptive. On Android platform, the average cost of each inference is about 120.80ms, and CPU and memory usage slightly increased 1.839\% and 2.06\% respectively. iOS platform has higher efficiency and lower resource consumption, with average cost of 49.39ms, and CPU and memory usage increased 0.488\% and 1.511\% respectively.
\begin{table}[h]
    \caption{Computing efficiency and resource usage of mobile ranking model with beam search on Android an iOS.} \label{tab:resource}
    \begin{tabular}{cccc}
    \toprule
    \textbf{Platform} & \textbf{Average Cost (ms)} & \textbf{CPU}  & \textbf{Memory} \\ \midrule
    Android & 120.80 &  +1.839\%  & +2.06\% \\ 
    iOS & 49.39 & +0.488\% & +1.511\% \\ \bottomrule
    \end{tabular}
\end{table}

\subsection{Case Study}\label{sec:case}

In this subsection, we show a representative case to visually demonstrate the effect and importance of real-time feedback.

As shown in \autoref{fig:case}, the user gave positive feedback to 5 out of 8 latest watched videos, including two videos in ``kids'' category (videos at position 72 and 74). When predicting user's preference on the candidate video from category ``kids'', we can see that the attention scores is higher on videos with explicit feedback and related category, showing that our model can successfully learn to attend to most related videos in watched history. For the candidate video, server-side predicted rate of like is merely 0.049, and client-side prediction is significantly higher at 0.592, which proves that user's real-time feedback have great influence on subsequent candidates. In the end, the user indeed liked this video and added it to favorite list. This case demonstrates that if we are able to make use of users' real-time feedback, we can understand their current interests much better.

\section{Conclusion}
In this paper, we propose a recommender framework on mobile devices for short video recommendation scenario, to solve the problem of untimely user real-time interest perception and content order adjustment. We specifically design a small model architecture that can be directly and completely deployed on mobile devices, which can make use of real-time user feedback to improve model prediction accuracy. Then we use a context-aware planning method to better capture the mutual influence between candidate videos, to recommend the video up-next. The whole framework is tested both offline and online in a billion-user scale short video application, and the result shows its superiority.

In the future, we will explore how to enhance the collaboration between recommender systems on mobile devices and in the cloud, in order to further improve user experience.
\bibliographystyle{ACM-Reference-Format}
\balance
\bibliography{main}
\end{document}